\documentclass[twocolumn,showpacs,preprintnumbers,amsmath,amssymb]{revtex4-1}
\usepackage{natbib}
\usepackage{graphicx}
\usepackage{dcolumn}
\usepackage[english]{babel}
\usepackage{amsmath}
\usepackage{amssymb}
\usepackage{tabularx}
\usepackage{multirow}
\usepackage[all]{xy}
\usepackage{appendix}
\usepackage{xspace}
\usepackage{color}
\newcommand{\ket}[1]{\ensuremath{|#1\rangle}\xspace}

\begin{document}
\title{Resonantly enhanced filamentation in gases}
\author{J. Doussot}
\author{G. Karras}
\author{F. Billard}
\author{P. B\'ejot} \email{pierre.bejot@u-bourgogne.fr}
\author{O. Faucher}
\affiliation{Laboratoire Interdisciplinaire CARNOT de Bourgogne, UMR 6303 CNRS-Universit\'e Bourgogne Franche-Comt\'e, 9 Av. A. Savary, BP 47870, F-21078 DIJON Cedex, France}


\begin{abstract}
In this Letter, a low-loss Kerr-driven optical filament in Krypton gas is experimentally reported in the ultraviolet. The experimental findings are supported by \textit{ab initio} quantum calculations describing the atomic optical response. Higher-order Kerr effect induced by three-photon resonant transitions is identified as the underlying physical mechanism responsible for the intensity stabilization during the filamentation process, while ionization plays only a minor role. This result goes beyond the commonly-admitted paradigm of filamentation, in which ionization is a necessary condition of the filament intensity clamping. At resonance, it is also experimentally demonstrated that the filament length is greatly extended because of a strong decrease of the optical losses.
\end{abstract}

\maketitle

Laser filamentation \cite{ChinReport,BergeReport,MysyReport} denotes the property of an ultrashort and high-power laser to self-organize in a very small structure able to transport very high intensities over distances far longer than those allowed by linear optics laws. The competition between the Kerr effect, responsible for beam self-focusing and plasma generation (i.e., ionization), responsible for beam defocusing, was rapidly identified as the paradigm describing the filamentation process. The unique self-guiding characteristic of a filament has been advantageously exploited in numerous remote applications \cite{KasparianW08}, in which the transport of high energies is of prime importance, such as nonlinear spectroscopy, remote sensing, and lightning protection, to cite a few.
 \begin{figure*}
	\includegraphics[width=17cm,keepaspectratio]{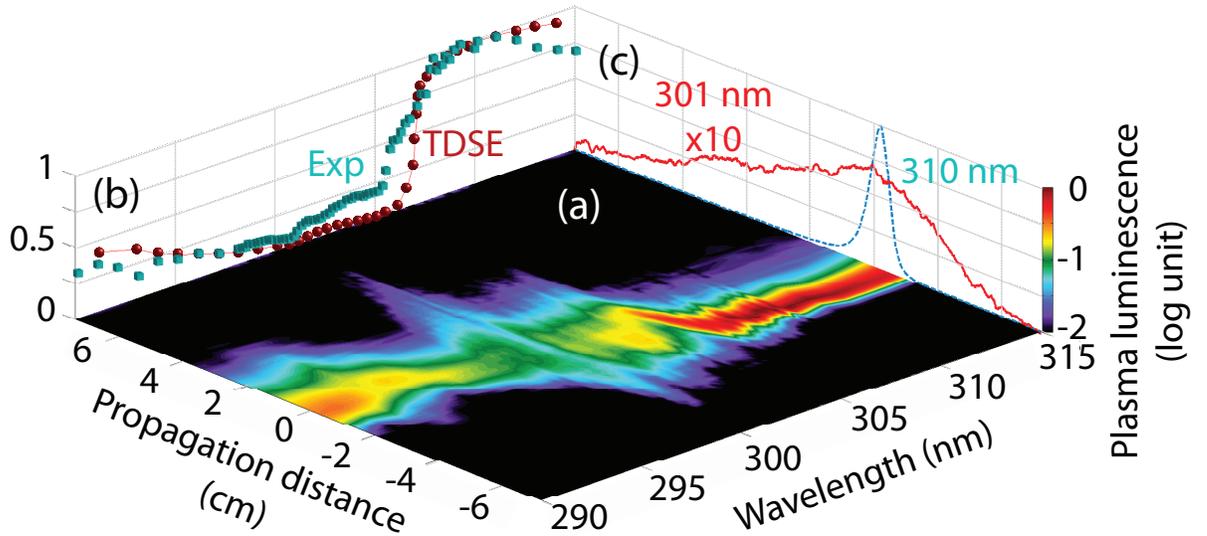}
	\caption{Longitudinal profile of the filament (a) and the associated maximal fluorescence signal (b) as a function of the central wavelength. The blue cubes in (b) corresponds to experimental results while the red spheres to \textit{ab initio} output calculations obtained in the case of a 50\,fs laser pulse. The panel (c) displays the longitudinal profile for two distinct wavelengths, namely 301\,nm and 310\,nm.}
	\label{FilamentProfile}
\end{figure*}
Ultimately, however, the finite energy contained in a filament is dissipated because of losses originated from ionization, limiting thereby the filament length. In other words, ionization represents a fundamental limitation in remote applications where long-ranged filaments are required. In order to circumvent this intrinsic restriction, methods to prolong these light strings have been suggested \cite{Polynkin,Fu,MoloRefuel}. All of them have been applied to 800\,nm laser pulses, mainly because almost all lasers able to supply enough power to trigger the filamentation process emit close to this wavelength. In fact, only a few works have studied the filamentation process in the case of laser operating at other wavelengths, mainly because of the lack of appropriate laser sources. Apart from 800\,nm, the filamentation process in gaseous media has been mainly studied in the case of lasers operating at harmonic frequencies of the Ti:sapphire lasers, i.e., at 400\,nm \cite{UVfilament1} or 266\,nm \cite{UVfilament2}. Some other experiments have also demonstrated filamentation in the case of a 3.9\,$\mu$m mid-infrared optical parametric chirped-pulse amplified laser \cite{Kartashov1,Kartashov2,Mitrofanov}, of a 248\,nm KrF laser \cite{Rambo}, and more recently of a 1.03\,$\mu$m picosecond laser \cite{MysyFilamentIR}. So far, optimizing the filamentation process by tuning the laser wavelength has been rather disregarded. In this Letter, it is experimentally shown that filamentation can be advantageously enhanced if the laser central wavelength is tuned close to a resonance with a multi-photon transition involving the ground and a bound excited state. More particularly, it is shown that tuning the central wavelength of the laser to a three-photon electronic resonance of krypton [$4s^24p^6$-$4s^24p^5(2P°_{3/2})6s$] at $\lambda$=300\,nm leads to a tenfold increase of the filament length accompanied together with a strong decrease of the nonlinear losses experienced by the filament together with a decrease by an order of magnitude of ionization. This experimental result is confirmed by \textit{ab-initio} calculations reproducing the optical response of the atom. Moreover, the theoretical calculations show that ionization is not responsible in this case for the intensity clamping of the filament, contrary to the theoretical paradigm generally describing the filamentation process. If transposed to molecular oxygen, involving for instance the $B^3\Sigma_u^-$-$X^3\Sigma_g^-$ ($v'=13$, $v"=0$) three-photon Schumann-Runge transition at 534\,nm, this result could find many atmospheric applications where long and low-loss filaments are needed.\\
\indent In the present experiment, filaments are produced in a 1.5\,m long cell filled with krypton at a 6\,bar pressure with a femtosecond tunable ultraviolet (UV) source focused with a $f=$1\,m spherical aluminium mirror. The UV pulse is produced by frequency quadrupling the output of a commercial non-collinear parametric amplifier operating in the 1100-2500\,nm spectral range. The latter is pumped by a 100\,fs 800\,nm 1\,kHz 2\,mJ chirped-pulse-amplified laser.
\begin{figure}[b!]
	\includegraphics[width=8.5cm,keepaspectratio]{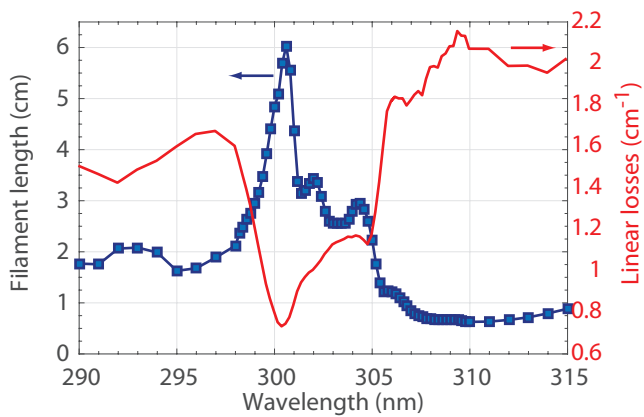}
	\caption{Filament length (blue squares) and losses per filament length unit (red solid line) as a function of the central wavelength of the filament.}
	\label{Losses}
\end{figure}
The energy of the UV pulse before the cell is set at 30\,$\mu$J in the whole wavelength range studied during the experiment (290-315\,nm). The pulse duration is estimated to be 50\,fs. The plasma fluorescence produced by the filament was imaged on the side of the cell with the help of a silicon camera coupled together with a commercial objective. A short-pass filter rejecting wavelengths longer than 675\,nm was placed on the objective in order to eventually suppress fluorescence signal induced by the bound excited state of krypton. The collected signal is then representative of ionization induced all along the propagation of the filament. Figure\,\ref{FilamentProfile}(a) shows the longitudinal profile (in log. unit) of the plasma fluorescence and Fig.\,\ref{FilamentProfile}(b) the maximum plasma fluorescence as a function of the laser central wavelength.
\begin{figure*}
	\includegraphics[width=17cm,keepaspectratio]{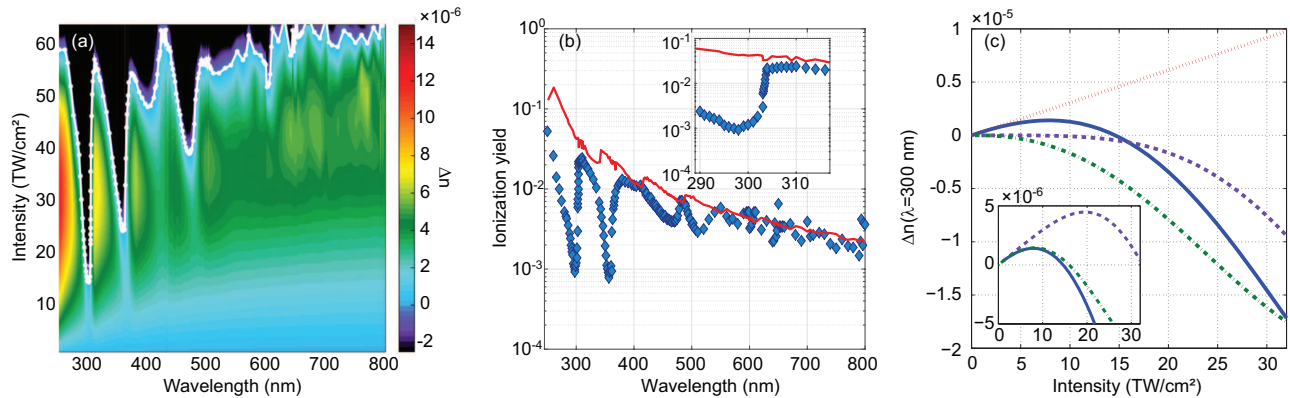}
	\caption{(a) Nonlinear refractive index of krypton gas as a function of both intensity and central wavelength of the 7\,fs laser field. The white line corresponds to the peak intensity $I_\textrm{inv}$ at which the nonlinear refractive index changes its sign. (b) Ionization yield as a function of the laser wavelength at the intensity at which the nonlinear refractive index changes its sign according to \textit{ab initio} calculations (blue diamonds) and according to the usual scenario of filamentation (red solid line). The inset corresponds to the wavelength region studied experimentally. (c) Contributions of the Kerr effect (dotted red), ionization (dashed violet) and higher-order Kerr effect (dotted-dashed green) to the total refractive index change (solid blue) of krypton at $\lambda$=300\,nm as a function of the intensity. The inset shows the total refractive index change (solid blue), the sum of the contributions of the Kerr effect and ionization (dashed violet) and the sum of the contributions of the Kerr effect and higher-order Kerr effect (dashed green).}
	\label{DNvsLambda}
\end{figure*}
Filaments produced with a pulse wavelength above 305\,nm (i.e., above the resonance) share in good approximation the same longitudinal profile and lead to a similar ionization yield. An abrupt change of both the longitudinal profile [see Fig.\,\ref{FilamentProfile}(c) that compares the longitudinal profile of filaments produced at 301\,nm and 310\,nm] and the plasma fluorescence amplitude takes place when the filament central wavelength is close or below the atomic resonance wavelength. As shown in Fig.\,\ref{Losses}, the filament length increases by a factor of ten when the central wavelength of the filament is tuned to the atomic resonance. This increase of the filament length is accompanied together with a decrease of ionization by an order of magnitude. Since multiphoton resonances are known to enhance, at a given intensity, the ionization process \cite{CormierIon}, this necessarily implies a strong decrease of the intensity within the filament close to the resonance. Moreover, the fact that longer filaments are obtained for a central wavelength minimizing ionization seems to corroborate the well-admitted scenario, in which ionization-induced absorption is the main mechanism limiting the filament length. In order to confirm this, the energy transmission has been measured as a function of the central wavelength of the filament. The associated optical losses experienced by the filament per unit length, i.e., the total optical losses normalized by the filament length are displayed in Fig.\,\ref{Losses}. Again, a rapid transition close to the atomic resonance is observed. More particularly, a strong reduction of the absorption is recorded around the resonance, which explains why the filament can sustain high intensity over longer distances in this spectral region. In order to better understand the underlying Physics, \textit{ab initio} calculations capturing the atomic optical response of krypton atoms have been performed. The simulations consist in solving the time-dependent Schr\"{o}dinger equation (TDSE) determining the temporal dynamics of the electronic wavepacket when submitted to a strong laser field. More specifically, under the single-active electron and dipole approximations, the three-dimensional TDSE describing the evolution of the electron wavefunction $\ket{\psi}$ in the presence of an electric field $\textbf{E}(t)$ reads:
\begin{equation}
i\frac{d\ket{\psi}}{dt}=(H_0+H_{\textrm{int}})\ket{\psi},
\end{equation}
where $H_0=\mathbf{\nabla}^2/2+V_{\textrm{eff}}$ is the atom Hamiltonian, $H_{\textrm{int}}=-\textbf{E}(t)\cdot\textbf{r}$ is the interaction term expressed in the length gauge. The effective potential $V_{\textrm{eff}}$ of Krypton used in the calculation accurately fits the atomic eigen-energies and -wavefunctions \cite{Cloux}:
	\begin{equation}
V_{\textrm{eff}}(r)=-\frac{1+A\textrm{e}^{-Br}+\left(35-A\right)\textrm{e}^{-Cr}}{r},
\label{PotEff}
	\end{equation}
with $A$=5.25, $B$=0.902, and $C$=3.64.
\begin{figure}
	\includegraphics[width=8.5cm,keepaspectratio]{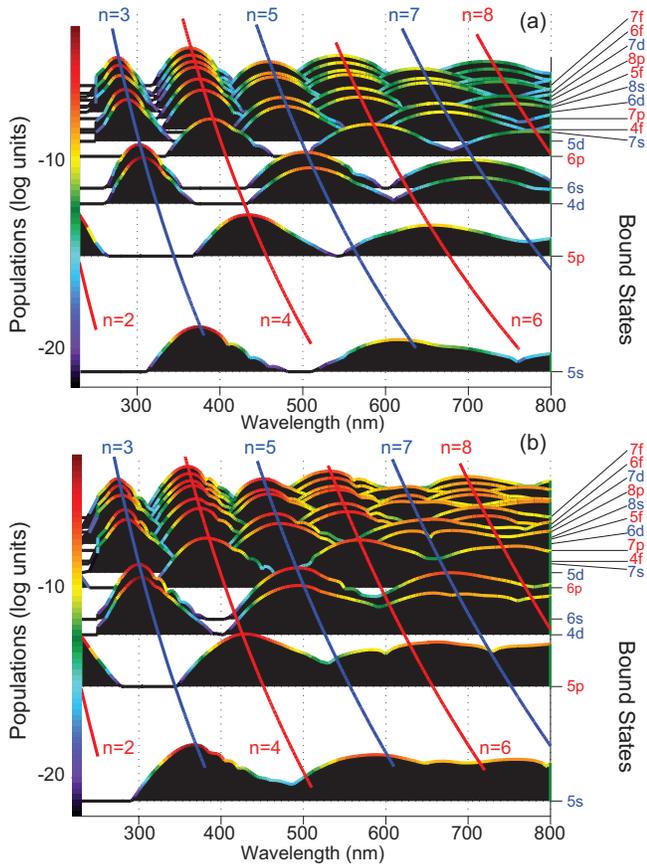}
	\caption{Populations left in the excited states as a function of the laser central wavelength at 2\,TW.cm$^{-2}$ (a) and 14\,TW.cm$^{-2}$ (b). The blue (resp. red) lines pinpoint the multiphoton resonances involving an odd (resp. even) number $n$ of photons. The 4p ground state is not shown for clarity. Note also that the displacement of these lines stemming from the dynamic Stark shift is taken into account in (b). The laser pulse duration is 7\,fs.}
	\label{Populations}
\end{figure}
Due to the deep potential well supporting inner shells, Eq.\,\ref{PotEff} cannot be used as such. To obtain a potential suitable for the intended simulations, a similar treatment as in \cite{MullerPotential} was performed. Inner shells can be eliminated without distorting the other wavefunctions by imposing an hard-core boundary condition $\psi$($R_0$)=0, with $R_0=0.615$\,a.u.. Finally, another modification of the potential was performed so as to eliminate the $4s$ state lying below the 4$p$ ground state and then ensure that no transition can take place between these states during the interaction. It was done by adding to the effective potential felt by the $s$ states the soft core potential $W(r)$:
	\begin{eqnarray}
\begin{aligned}
W(r<R_x)&=G&\left(\frac{1}{r}-\frac{1}{R_x}\right)^2,\\
W(r>R_x)&=0,& \nonumber
\end{aligned}
	\end{eqnarray}
with $G=50$ and $R_x$=3.35. The atom is initially in the ground state (4p) and the electric field $E$ is linearly polarized along the $z$ axis and has a gaussian temporal envelope. The pulse duration is kept constant when changing the central wavelength. The calculations have been performed for 7\,fs and 50\,fs intensity pulse durations (at full-width at half-maximum). The first set of calculations (7\,fs) have been conducted over a broad spectral range (250-800\,nm) while the second one (50\,fs) was performed on the spectral region experimentally studied (290-315\,nm). It was checked that the use of different pulse durations does not change the qualitative conclusion of the numerical work. As already described in \cite{BejotPRL3}, the knowledge of the electronic wavefunction all along the interaction between the atom and the field then allows to evaluate all physical parameters of interest, namely the nonlinear refractive index, the population left after the interaction in the ground and the bound excited states, and the ionization yield. Figure \ref{DNvsLambda}(a) displays the nonlinear refractive index as a function of both the central wavelength and the peak intensity of the 7\,fs laser pulse. The same behavior is noticed for every wavelengths as the intensity is increased. The nonlinear refractive index first increases almost linearly with respect to the intensity, which corresponds to the well-known Kerr effect. It then saturates and finally becomes negative as the intensity reaches a few tens of terawatts per square centimeters. The intensity $I_{\textrm{inv}}$ at which the refractive index becomes negative [white line in Fig.\ref{DNvsLambda}(a)] is almost the same between 500\,nm and 800\,nm and lies at about 60\,TW.cm$^{-2}$. This indicates that the clamping is induced by an off-resonant process and that the exact atomic structure of the atom plays no role. Below 500\,nm, however, \textit{ab initio} calculations foresee that $I_{\textrm{inv}}$ strongly decreases in three different localized wavelength regions, namely around 470\,nm, 357\,nm, and 300\,nm. More particularly, at these wavelengths, the clamping intensity drops to 39, 24, and 14\,TW.cm$^{-2}$, respectively. In parallel, the ionization level obtained at the clamping intensity strongly drops by an order of magnitude. The expected decrease of ionization is in excellent quantitative agreement with the experimental observations as shown in Fig.\,\ref{FilamentProfile}(b). The decrease of the ionization yield, noticed in both our experimental and theoretical results, goes beyond the common scenario of filamentation, in which the intensity clamping is due to ionization. In this paradigm, the stabilization of the filamentation process occurs when $n_2I=\frac{\rho}{2\rho_\textrm{c}}$, where $n_2$ is the nonlinear refractive index of the medium, $\rho_\textrm{c}=\epsilon_0m_\textrm{e}\omega_0^2/q^2$ is the critical plasma density, $I$ is the laser intensity, $\rho$ the ionization yield, $\epsilon_0$ is the vacuum permittivity, $m_\textrm{e}$ and $q$ are the electron mass and charge, and $\omega_0$ is the laser pulsation. The red line in Fig.\,\ref{DNvsLambda}(b) shows the ionization yield at which the nonlinear refractive index changes its sign calculated in this framework. In very good agreement with TDSE calculations performed in the infrared region, the usual scenario fails to describe the strong decrease of ionization taking place in the ultraviolet and blue regions. More particularly, the calculations reveal that the ionization contributes to only 10\,\% to the intensity clamping at $\lambda=300$\,nm and consequently is not the mechanism responsible for the intensity stabilization occurring during the filamentation process in this spectral domain [see Fig\,\ref{DNvsLambda}(c)]. This observation is in complete agreement with the experimental findings shown above and confirms the existence of a Kerr-driven filamentation regime. In the present work, the clamping mechanism is due to resonant three-photon transitions leading to a giant defocusing resulting from higher-order Kerr effects. It is analogous to the sign inversion of the nonlinear refractive index $n_2$ occurring close to a two-photon resonance \cite{BergeDefoc}. As we see here, the same phenomenon takes place close to a three-photon transition for the higher-order nonlinear refractive index $n_4$. The latter is estimated from \textit{ab initio} calculations to $n_4=-5.03\,10^{-8}$\,cm$^4$.TW$^{-2}$ at $\lambda$=300\,nm. In order to illustrate this resonant phenomenon, the populations left in the excited states as a function of the central wavelength is depicted in Fig.\,\ref{Populations} for 2\,TW.cm$^{-2}$ (a) and 14\,TW.cm$^{-2}$ (b) peak field intensities. At low intensity, the dynamic Stark shift is negligible so that the eigen-energies of the excited levels are not perturbed by the field. As a consequence, the bound states are populated significantly through an allowed multiphoton transition. At higher peak intensity, the dynamic Stark shift starts to play a role and displaces the wavelengths at which resonances take place. This is especially the case for longer wavelengths since the energy shift varies as $\lambda^2$ \cite{Kaminski}. As a consequence, while sharp resonances still take place in the ultraviolet at higher intensities, they are completely smeared out in the infrared region. This explains why resonance-induced effects can only be observed in the ultraviolet. A close look at the population confirms that a three-photon resonant transition takes place at 300\,nm, namely the 4p-6s transition. More particularly, two three-photons resonances take place at this wavelength because of the finite spectral extend of the laser field: the 4p-6s and 4p-4d transitions.\\
\indent As a conclusion, the existence of a Kerr-driven regime of filamentation is experimentally demonstrated in krypton at 300\,nm. This finding contrasts with the conventional model describing the filamentation process, in which ionization drives the filament stabilization. The experimental findings are supported by \textit{ab initio} quantum calculations which are in excellent agreement with the experiments. The theoretical results pinpoint the role of three-photon resonant transitions in this Kerr-driven ultraviolet filamentation. It is also experimentally demonstrated that the filament length is greatly extended at resonance due to a strong decrease of the optical losses. This work could be extended to molecular gases of atmospheric interest, such as oxygen and nitrogen, where it would lead to numerous applications. For instance, the three-photon Schumann-Runge transition $B^3\Sigma_u^-$-$X^3\Sigma_g^-$ ($v'=13$, $v"=0$) could be resonantly excited using a laser at 534\,nm. The generalization of the present work to molecular gases could motivate the development of intense laser sources emitting in the visible region with goal to optimize the filamentation process and its underlying atmospheric applications.
\acknowledgments
This work was supported by the Conseil R\'egional de Bourgogne (PARI program), the CNRS, the French National Research Agency (ANR)
through the CoConicS program (contract ANR-13-BS08-0013) and the Labex ACTION program (contract ANR-11-LABX-0001-01).
P.B. thanks the CRI-CCUB for CPU loan on its multiprocessor server and B. Lavorel for fruitful discussions.
\bibliographystyle{srt}

\end{document}